\documentclass[11pt]{article}

\usepackage{amsmath,amssymb}
\usepackage[active]{srcltx}
\usepackage{hyperref}
\usepackage{xcolor}
\def\tabaddress#1{{\small\it\begin{tabular}[t]{c}#1
\\[1.2ex]\end{tabular}}}
\def\UPCMAT{Departamento de Matem\'aticas.
   Campus Norte U.P.C., Ed. C-3\\
   C/ Jordi Girona 1.
   E-08034 Barcelona, SPAIN}

\font\fr=eufm10 scaled \magstep 1 
\newtheorem{teor}{Theorem}
\newtheorem{prop}{Proposition}

\newtheorem{definition}{Definition}

\def\beq{\begin{equation}}
\def\eeq{\end{equation}}
\def\bea{\begin{eqnarray}}
\def\eea{\end{eqnarray}}
\def\beann{\begin{eqnarray*}}
\def\eeann{\end{eqnarray*}}
\def\ben{\begin{enumerate}}
\def\een{\end{enumerate}}
\def\bit{\begin{itemize}}
\def\eit{\end{itemize}}

\def\moment#1#2#3{{#1}_{#2}, \ldots, {#1}_{#3}}

\def\qed{\ifvmode\removelastskip\fi
{\unskip\nobreak\hfil\penalty50\hbox{}\nobreak\hfil \hbox{\vrule
height1.2ex width1.2ex}\parfillskip=0pt \finalhyphendemerits=0
\par\smallskip}}
\def\vf{\mbox{\fr X}}
\def\df{{\mit\Omega}}
\def\Lag{{\cal L}}
\def\lag{\pounds}

\def\d{{\rm d}}
\def\Nat{\mathbb{N}}
 
\def\Real{\mathbb{R}}

\def\inn{\mathop{i}\nolimits}
\def\Tan{{\rm T}}

\def\Lie{\mathop{\rm L}\nolimits}
\def\ls{(J^1\pi,\Omega_\Lag )}

\def\Cinfty{{\rm C}^\infty}
\def\proof{( {\sl Proof} )\quad}

\title{REMARKS ON MULTISYMPLECTIC REDUCTION}
\author{\sc Arturo Echeverr\'{\i}a-Enr\'\i quez,
 Miguel C. Mu\~noz-Lecanda,
\\
\sc Narciso Rom\'an-Roy
 \\
   \tabaddress{\UPCMAT}\\
{\small ({\bf e}-{\it mails}: miguel.carlos.munoz@upc.edu,
narciso.roman@upc.edu)}}

\begin{document}

\maketitle

\begin{abstract}
The problem of reduction of multisymplectic
manifolds by the action of Lie groups is stated and discussed,
as a previous step to give a fully covariant scheme
of reduction for classical field theories with symmetries.
\end{abstract}

\begin{small}
\begin{center}

AMS s.\,c.\,(2010): 53D05, 53D20, 55R10, 57M60, 57S25, 70S05, 70S10.
\end{center}
\end{small}

\section{Introduction}

Multisymplectic manifolds constitute one of the most used 
and generic geometric frameworks for classical field theories.
Then, the covariant reduction of field theories by symmetries requires,
as a previous step, to study the reduction of multisymplectic manifolds. 

This procedure should be based on the pioneering
Marsden-Weinstein technique of reduction of symplectic manifolds \cite{MW-rsms} which was applied later to many different situations such as, for instance,
the reduction of autonomous and non-autonomous Hamiltonian and
Lagrangian systems (regular or singular)
\cite{Al-89,ACG-91,CCCI-86,LMR-92,LS-93,EMR-99,IM-92,SL-91},
non-holonomic systems \cite{BS-93,BKMM-1996,CLMM-98,Ma-95},
control systems \cite{BC-99,EMMR-2003,Ma-2004,MCL-01,Su-95,Sh-87},
and in other cases (Poisson, Dirac, Euler-Poincar\'e, Routh and implicit reduction)
\cite{BVS-2000,LM-95,MR-86,MRW-84,MS-93}.

In relation to the problem of reduction of classical field theories,
only partial results have been achieved in the context of
Lagrangian, Poisson and Euler-Poincar\'e reduction, and other particular situations in multisymplectic field theories
\cite{BuCaIg,CGR-2007,CM-2003,CR-2003,FGR-2001,MaSw1,Sn-2004,Va-2007}.
Nevertheless, the problem of establishing a complete scheme of reduction 
for the multisymplectic case (in the line of the
Marsden-Weinstein theorem), which should give a fully 
covariant reduction of the theory, is still unsolved.

The aim of this letter is to review the statement of the problem and
give some insights in this way.
In particular, a brief discussion about Noether invariants in Lagrangian field theory illustrates these considerations.

In this paper, manifolds are real, paracompact,
connected and $C^\infty$, and maps are $C^\infty$.

\section{Multisymplectic manifolds. Actions of Lie groups}

Let ${\cal M}$ be a $m$-dimensional differentiable manifold, and
$\Omega\in\df^{k+1}({\cal M})$ a differentiable form in ${\cal M}$ 
($k+1\leq m$).
For every $x\in {\cal M}$, the form $\Omega_x$ eestablish a correspondence
$\hat{\Omega}_r(x)$ between the set of 
$r$-vectors, $\Lambda^r\Tan_x {\cal M}$, and the set of $(k+1-r)$-forms,
$\Lambda^{k+1-r}\Tan_x^* {\cal M}$, as
$$
\hat\Omega_r(x) \colon\Lambda^r\Tan_x {\cal M} 
\longrightarrow \Lambda^{k+1-r}\Tan_x^* {\cal M}
\quad ; \quad
 v  \mapsto \inn(v)\Omega_x \ .
$$ 
If $v$ is homogeneous,
$v = v_1\wedge\ldots\wedge v_r$, then $\inn (v)\Omega_x = \inn
(v_1\wedge\ldots\wedge v_r)\Omega_x = 
\inn (v_1)\ldots\inn (v_r)\Omega_x$.
Thus, an $r$-vector field $X\in\vf^r({\cal M})$ 
(that is, a section of $\Lambda^r\Tan {\cal M}$)
defines a contraction 
$\inn (X)$ of degree $r$ of the algebra of differential forms in ${\cal M}$.
The $(k+1)$-form $\Omega$ is {\sl $1$-nondegenerate} if
$\ker\Omega:=\{ X\in\vf ({\cal M})\, \vert\, \hat\Omega_1(x)(X)=0;\
\mbox{\rm  $x\in {\cal M}$}\}=\{ 0\}$.

A couple $({\cal M},\Omega)$ is a
{\sl multisymplectic manifold} if $\Omega$ is closed and 
$1$-nondegenerate.
The degree $k+1$ of the form
$\Omega$ will be called {\sl the degree of the multisymplectic manifold}.
$X\in\vf({\cal M})$ is a {\sl Hamiltonian vector field}
if $\inn (X)\Omega$ is an exact $k$-form; that is,
there exists $\zeta \in\df^{k-1}({\cal M})$ such that
\beq
\inn (X)\Omega =\d\zeta \ .
\label{ham}
\eeq
$\zeta$ is defined modulo closed $(k-1)$-forms.  The class
$\bar\zeta\in\df^{k-1}({\cal M})/Z^{k-1}({\cal M})$ 
defined by $\zeta$ is called the
{\sl Hamiltonian} for $X$, and every element in this class 
is a {\sl Hamiltonian form} for $X$.
Furthermore,
$X\in\vf({\cal M})$ is a {\sl locally Hamiltonian vector field}
if $\inn (X)\Omega$ is a closed $k$-form.
Then, for every $x\in {\cal M}$, there is an open
neighbourhood $W\subset {\cal M}$ and $\zeta \in\df^{k-1}(W)$ such that (\ref{ham}) holds on $W$.
As above, changing ${\cal M}$ by $W$, we obtain the
{\sl Hamiltonian} for $X$, $\bar\zeta\in\df^{k-1}(W)/Z^{k-1}(W)$,
and the {\rm local Hamiltonian forms} for $X$.

Conversely, $\zeta \in\df^{k-1}({\cal M})$ (resp. $\zeta \in\df^{k-1}(W)$)
is a {\sl Hamiltonian form} (resp. 
a {\sl local Hamiltonian form}) if there exist a 
vector field $X_\zeta\in\vf({\cal M})$ (resp. $X_\zeta\in\vf({\cal M})$) 
such that (\ref{ham}) holds (resp. on $W$).
Of course, a vector field $X\in\vf({\cal M})$ is a locally Hamiltonian
vector field if, and only if, the Lie derivative $\Lie(X)\Omega =0$. 
If $X,Y$ are locally Hamiltonian vector fields, 
then $[X,Y]$ is a Hamiltonian vector field with Hamiltonian form 
$\inn(X\wedge Y)\Omega$.

We denote by ${\cal H}^{k-1}({\cal M})$ the $\Real$-vector space of
Hamiltonian $(k-1)$-forms for the $(k+1)$-multisymplectic form $\Omega$,
and
${\bar{\mathcal H}}^{k-1}({\cal M})$ the set of
Hamiltonian $(k-1)$-forms modulo closed
$(k-1)$-forms; that is,  ${\bar{\mathcal H}}^{k-1}({\cal M}) = {\mathcal H}^{k-1}({\cal M})/Z^{k-1}({\cal M})$. The classes in
${\bar{\mathcal H}}^{k-1}(M)$ are denoted by
$\bar{\zeta}$ (this is the class of which the Hamiltonian
 $(k-1)$-form $\zeta$ is a representative).
Then there is a natural Lie algebra
structure on  ${\bar{\mathcal H}}^{k-1}(M)$ defined as follows (see \cite{Ca96a,IEMR-2012}):
given $\xi,\zeta\in\tilde{\mathcal H}^{k-1}({\cal M})$, 
let $X_{\xi},X_{\zeta}\in\vf({\cal M})$ be
their corresponding Hamiltonian vector fields;
the {\sl bracket of the Hamiltonian classes} $\bar\xi,\bar\zeta\in\bar{\mathcal H}^{k-1}({\cal M})$
is the class $\{\bar\xi ,\bar\zeta\}\in\tilde{\mathcal H}^{k-1}({\cal M})$ of which the Hamiltonian
 $(k-1)$-form
$$
\{\xi,\zeta\}:=-\inn(X_\xi)\inn(X_\zeta)\Omega
$$
is a representative. We commit an abuse of notation and denote $\{\bar\xi ,\bar\zeta\}$ simply as $\{\xi,\zeta\}$.

\begin{definition}
Let $\Phi \colon G\times {\cal M} \to {\cal M}$ be
an action of a Lie group $G$ on a multisymplectic manifold
$({\cal M},\Omega )$. We say that
$\Phi$ is a {\rm multisymplectic action}
(or also that $G$ {\rm acts multisymplectically} on ${\cal M}$ by $\Phi$)
if, for every $g \in G$, $\Phi_g$ is a multisymplectomorphism,
that is, $\Phi_g^*\Omega = \Omega$.
Then ${\cal M}$ is a {\rm multisymplectic $G$-space},
or also that $G$ is a {\rm symmetry group} of $({\cal M},\Omega)$.
\end{definition}

We denote by $\tilde{\mathfrak g}\subset\vf ({\cal M})$ the real Lie algebra of 
fundamental vector fields.
As a consequence of the definition, the fundamental vector field
$\tilde\xi\in\tilde{\mathfrak g}$
associated with every $\xi\in{\mathfrak g}$ by $\Phi$ is a
locally Hamiltonian vector field,
$\tilde\xi\in \vf_{{\rm lH}}({\cal M})$
(conversely, if for every $\tilde\xi\in{\mathfrak g}$, we have that
$\tilde\xi\in \vf_{{\rm lH}}({\cal M})$, then 
$\Phi$ is a multisymplectic action of $G$ on ${\cal M}$).
So we have that,
for every $\xi\in {\mathfrak g}$, $\Lie(\tilde\xi )\Omega = 0$ or,
what is equivalent, $\inn(\tilde\xi)\Omega \in Z^k({\cal M})$
(it is a closed $k$-form). Then,
following the same terminology as for actions
of Lie groups on symplectic manifolds
\cite{AM-78}, \cite{LM-87}, \cite{Wa-71},
we define:

\begin{definition}
Let $\Phi \colon G\times {\cal M}\to {\cal M}$ be a multisymplectic action 
of $G$ on ${\cal M}$. $\Phi$ is a
{\rm strongly multisymplectic} or {\rm Hamiltonian action}
 if $\tilde{\mathfrak g}\subseteq\vf_{\rm H}({\cal M})$
or, what is equivalent, for every $\xi\in {\mathfrak g}$,
$\inn(\tilde\xi)\Omega$ is an exact form.
Otherwise, it is called a {\rm locally Hamiltonian action}.
\end{definition}

In particular, if $({\cal M},\Omega )$ is an exact multisymplectic manifold 
(that is, there exists $\Theta\in\df^k({\cal M})$
such that $\d \Theta = \Omega$), and $\Phi$ an {\rm exact action};
that is, $\Phi_g^*\Theta = \Theta$, for every $g \in G$,
then $\tilde{\mathfrak g}\subset\vf_{\rm H}({\cal M})$ and,
for every $\tilde\xi\in\tilde{\mathfrak g}$, 
its Hamiltonian form is $\zeta_\xi=-\inn (\tilde\xi)\Theta$.
Hence, $\Phi$ is strongly multisymplectic.

\section{Momentum map for multisymplectic actions}

From now on $({\cal M},\Omega )$ will be a
$m$-dimensional multisymplectic manifold of degree $k+1$, and
$\Phi \colon G\times {\cal M} \to {\cal M}$ a strongly multisymplectic action of 
a Lie group $G$ on ${\cal M}$, with $\dim\,G=n$.

\begin{definition}
A {\rm comomentum map} associated with $\Phi$
is a map ${\cal J}^*\colon{\mathfrak g}\to{\cal H}^{k-1}({\cal M})$,
such that 
$$
\inn(\tilde\xi)\Omega=\d{\cal J}^*(\xi)\quad ;\quad \xi\in{\mathfrak g} \ .
$$
A {\rm momentum map} associated with $\Phi$
is a map
${\cal J}\colon {\cal M}\to{\mathfrak g}^*\otimes_{\cal M}\Lambda^{k-1}\Tan^*{\cal M}$
such that
$$
{\cal J}(x)(\xi):={\cal J}^*(\xi)(x)\in\Lambda^{k-1}\Tan_x^*{\cal M}
\quad ; \quad x\in {\cal M}\, ,\, \xi\in{\mathfrak g} \ .
$$
\label{mmmap}
\end{definition}

The comomentum maps are parametrized by
$L_{\Real}({\mathfrak g},Z^{k-1}({\cal M}))$;
that is, the real space of the linear maps from ${\mathfrak g}$
to $Z^{k-1}({\cal M})$.
In fact, if ${\cal F}\colon {\mathfrak g} \to Z^{k-1}({\cal M})$
is a continuous linear map,
and ${\cal J}^*$ is a comomentum map,
so is ${\cal J}^{'*}={\cal J}^*+{\cal F}$.
Furthermore,
for all $\xi_1,\xi_2\in {\mathfrak g}$, with
$\inn(\tilde\xi_i)\Omega=\d\zeta_{\xi_i}$, ($i=1,2$), then
$\inn([\tilde\xi_1,\tilde\xi_2])\Omega=
\d\zeta_{[\tilde\xi_2,\tilde\xi_1]}$,
and we have that
$$
\d\{\zeta_{\xi_1},\zeta_{\xi_2}\} =
\inn([\tilde\xi_1,\tilde\xi_2])\Omega=
\d\zeta_{[\xi_2,\xi_1]} \ ,
$$
therefore
 $\d\{\zeta_{\xi_1},\zeta_{\xi_2}\}=\d\zeta_{[\xi_2,\xi_1]}$,
and
$\{\zeta_{\xi_1},\zeta_{\xi_2}\} =
\zeta_{[\xi_2,\xi_1]}+\gamma (\tilde\xi_1,\tilde\xi_2)$,
where 
$\gamma\colon\tilde{\mathfrak g}\times\tilde{\mathfrak g}\to Z^{k-1}({\cal M})$
is~a skewsymmetric bilinear map.
Then the comomentum map is a Lie algebra homomorphism
if, and only if, $\gamma=0$.

\begin{definition}
$\Phi$ is a {\rm Poissonian} 
or {\rm strongly Hamiltonian action} if
there exists a comomentum map which is a Lie algebra homomorphism;
that is,
$\{{\cal J}^*(\xi_1),{\cal J}^*(\xi_2)\}={\cal J}^*([\xi_1,\xi_2])$,
for every $\xi_1,\xi_2\in{\mathfrak g}$.

$\Phi$ is a {\rm Coad-equivariant action} if
there exists a momentum map which is $Ad^*$-equivariant; that is,
for every $g\in G$, we have the following commutative diagram:
$$
\begin{array}{ccccc}
& & {\cal J} & & \\
&{\cal M}&\longrightarrow&{\mathfrak g}^*\otimes\Lambda^{k-1}\Tan^*{\cal M}& \\
\Phi_g &\Big\downarrow& &\Big\downarrow& 
{\rm Ad}_g^*\otimes\Lambda^{k-1}\Tan^*\Phi_{g^{-1}} \\
&{\cal M}&\longrightarrow&{\mathfrak g}^*\otimes\Lambda^{k-1}\Tan^*{\cal M}& \\
& & {\cal J} & &
\end{array}
$$
\end{definition}

Using the same reasoning than for actions of Lie groups on
symplectic manifolds \cite{AM-78}, one can prove that every
Coad-equivariant action is Poissonian.
As a particular case we have:

\begin{prop}
If $({\cal M},\Omega)$ is an exact multisymplectic manifold
with $\Omega = \d\Theta$,
and the action $\Phi$ is exact, then
a comomentum map exists which is given by
${\cal J}^*(\xi)=-\inn (\tilde\xi )\Theta$,
$\xi\in {\mathfrak g}$,
and the action is Coad-equivariant and Poissonian.
\end{prop}
\proof
In fact, for every $x\in {\cal M}$, $\xi\in{\mathfrak g}$, 
and $\moment{X}{1}{k-1}\in\Tan_x{\cal M}$, we have that
\beann
({\cal J}\circ\Phi_g)(x)(\xi ;\moment{X}{1}{k-1})&=&
{\cal J}(\Phi_g(x))(\xi ;\moment{X}{1}{k-1})=
{\cal J}^*(\xi)(\Phi_g(x) ;\moment{X}{1}{k-1})
\\ &=&
 -[\inn (\tilde\xi )\Theta](\Phi_g(x) ;\moment{X}{1}{k-1})
\\ &=&
-[\Phi_g^*\inn (\tilde\xi )\Theta]
(x ;\moment{\Tan_{\Phi_g(x)}\Phi_{g^{-1}}X}{1}{k-1}) \ ;
\eeann
but, bearing in mind that
$\widetilde{({\rm Ad}_g\xi)}=\Phi_{{g^{-1}}*}\tilde\xi$, we have
$$
\Phi_g^*\inn (\tilde\xi)\Theta=
\inn(\Phi_{{g^{-1}}*}\tilde\xi)\Phi_g^*\Theta=
\inn(\Phi_{{g^{-1}}*}\tilde\xi)\Theta=
\inn\widetilde{({\rm Ad}_g\xi)}\Theta
\ \Longrightarrow \ 
{\cal J}^*({\rm Ad}_g\xi)=\Phi_g^*\inn (\tilde\xi)\Theta=
-\Phi_g^*({\cal J}^*(\xi)) \ ,
$$
and hence
$$
{\cal J}(x)({\rm Ad}_g\xi;\moment{X}{1}{k-1})=
{\cal J}^*({\rm Ad}_g\xi) (x;\moment{X}{1}{k-1})=
-[\Phi_g^*\inn (\tilde\xi)\Theta](x;\moment{X}{1}{k-1}) \ .
$$
Therefore
\beann
({\cal J}\circ\Phi_g)(x)(\xi ;\moment{X}{1}{k-1})&=&{\cal J}(x)
({\rm Ad}_g\xi;\moment{\Tan_{\Phi_g(x)}\Phi_{g^{-1}}X}{1}{k-1})
\\ &=&
{\cal J}(x)
[({\rm Ad}_g\otimes\Lambda^{k-1}\Tan\Phi_{g^{-1}})
(\xi;\moment{X}{1}{k-1})]
\\ &=&
[({\rm Ad}_g\otimes\Lambda^{k-1}\Tan\Phi_{g^{-1}})^t({\cal J}(x))]
(\xi;\moment{X}{1}{k-1})
\\ &=&
[({\rm Ad}_g^*\otimes\Lambda^{k-1}\Tan\Phi_{g^{-1}})\circ{\cal J}](x)
(\xi;\moment{X}{1}{k-1}) \ ;
\eeann
thus the action is Coad-equivariant. Hence it is Poissonian too, and
$$
\zeta_{[\xi_1,\xi_2]}=-\Theta ([\tilde\xi_1,\tilde\xi_2])=
-\inn([\tilde\xi_1,\tilde\xi_2])\Theta =
-\Lie(\tilde\xi_1)\inn(\tilde\xi_2)\Theta = 
\{\zeta_{\xi_1},\zeta_{\xi_2}\} \ .
$$
\qed

\section{Momentum-type submanifolds and multisymplectic reduction}

\begin{definition}
A submanifold $S$ of ${\cal M}$,
with natural embedding  $j_S\colon S\hookrightarrow {\cal M}$, is
a {\rm momentum-type submanifold} if:
\begin{enumerate}
\item
$S$ is a closed submanifold of ${\cal M}$.
\item 
$j_S^*\inn(\tilde\xi)\Omega=0$, for every $\xi\in{\mathfrak g}$;
that is, $S$ is an integral submanifold of the exterior differential system
$\{\inn(\tilde\xi)\Omega\, ; \, \xi\in{\mathfrak g}\}$.
\item
$S$ is maximal, in the order eestablished by the inclusion,
among all the submanifolds verifying the above conditions.
\end{enumerate} 
\label{defin}
\end{definition}

Let $G_S\subset G$ be the maximal subgroup of $G$
(with respect to the inclusion) that leaves $S$ invariant
(the isotropy group of $S$).

Consider the submodule ${\cal G}\subset\vf({\cal M})$ defined by
${\cal G}:=\Cinfty({\cal M})\otimes_{\Real}\tilde{\mathfrak g}$.
This module generates the distribution
tangent to the orbits of the action of $G$ on ${\cal M}$; that is,
if $G_p$ denotes the orbit passing through $p\in {\cal M}$,
and $\langle\xi_1,\ldots,\xi_n\rangle$
is a basis of ${\mathfrak g}$, then 
$\Tan_pG_p=\langle(\tilde\xi_1)_p,\ldots,(\tilde\xi_n)_p\rangle
\subset\Tan_p{\cal M}$.

If $S$ is a closed submanifold of ${\cal M}$,
and ${\mathfrak g}_S\subset{\mathfrak g}$ is the Lie subalgebra
associated with $G_S$, we define
${\cal G}_S:=\Cinfty({\cal M})\otimes_{\Real}\tilde{\mathfrak g}_S$.
Then, for every momentum-type submanifold $S$ we have that
${\cal G}_S$ is a submodule closed under the Lie bracket.
Observe that, for every $\tilde\xi\in{\cal G}_S$,
as $\tilde\xi$ is tangent to $S$, 
there exists $\tilde\xi_S\in\vf(S)$ such that
$j_{S*}\tilde\xi_S=\tilde\xi\vert_S$,
and we have that
\beq
j_S^*[\inn(\tilde\xi)\Omega]=\inn(\tilde\xi_S)(j_S^*\Omega)=0 \ .
\label{GHpropo}
\eeq

Consider the ideal
$Nul(S):=\{ f\in\Cinfty({\cal M})\,\vert\, j_S^*f=0\}$,
and let $\underline{\vf(S)}$ be the set of vector fields in ${\cal M}$ 
which are tangent to $S$, that is, 
$\underline{\vf(S)}:=
\{ X\in\vf({\cal M})\,\vert\, \Lie(X)(Nul(S))\subset Nul(S)\}$.
Then we have the following obvious result:

\begin{prop}
Let $S$ be a closed submanifold of ${\cal M}$
with $\dim\, S\geq k$, and let $\underline{\vf^k(S)}$ be
the set of $k$-vector fields in ${\cal M}$ which are tangent to $S$.
Then $S$ is an integral submanifold of the exterior differential system
$\{\inn(\tilde\xi)\Omega\, ; \, \xi\in{\mathfrak g}\}$
if, and only if,
$$
\Cinfty({\cal M})\otimes_{\Real}
\{\inn(\tilde\xi)\Omega\, ; \, \xi\in{\mathfrak g}\}\subset(\underline{\vf^k(S)})' \ ,
$$
where 
$(\underline{\vf^k(S)})':=\{\alpha\in\df^k({\cal M})\,\vert\, 
\inn(X)\alpha=0\, ,\, \mbox{\rm  $X\in\underline{\vf^k(S)}$}\}$ is
the annihilator of $\underline{\vf^k(S)}$.
\end{prop}

If $\dim\, S=s\geq k$,
as $\displaystyle\dim\,\vf^k(S)={{s}\choose{k}}$,
and $\displaystyle\dim\,(\vf^k(S))'=
{{m}\choose{k}}-{{s}\choose{k}}$,
then, the condition for $S$ to be an integral submanifold of the 
above exterior differential system is that
\beq
n\leq {{m}\choose{k}}-{{s}\choose{k}}
\quad ; \quad
\mbox{\rm that is\ ,}\quad
{{s}\choose{k}}\leq {{m}\choose{k}}-n \ ,
\label{dim}
\eeq
(which is a condition on $s=\dim\, S$).
In particular,
\beq
{{s}\choose{k}}={{m}\choose{k}}-n
\quad \Longleftrightarrow\quad
\Cinfty({\cal M})\otimes_{\Real}
\{\inn(\tilde\xi)\Omega\, ; \, \xi\in{\mathfrak g}\}=(\underline{\vf^k(S)})' \ .
\label{optimal}
\eeq

\begin{definition}
Let $S$ be a momentum-type submanifold.
$S$ will be called an {\rm optimal momentum-type submanifold} if
$$
s\equiv\dim\, S= {\rm sup}\,\left\{ q\in\Nat\, \mid\,
{{q}\choose{k}}\leq {{m}\choose{k}}-n\right\} \ .
$$
In particular, $S$ will be called a {\rm maximal momentum-type submanifold} if
(\ref{optimal}) holds.
\end{definition}

Let $j_S\colon S\hookrightarrow {\cal M}$ 
be a momentum-type submanifold, with $\dim\, S\geq k+1$.
Denote $\omega=j_S^*\Omega\in\df^{k+1}(S)$,
which is a closed but, in general, $1$-degenerate form,
and let ${\cal K}_\omega$ be the distribution
associated with $\ker\,\omega$ 
(the characteristic distribution of $\omega$).
We have that ${\cal G}_S$ is an involutive distribution
(also denoted by ${\cal G}_S$) which,
as a consequence of (\ref{GHpropo}),
is a subbundle of ${\cal K}_\omega$. Therefore:

\begin{teor}
If the action of $G_S$ on $S$ is free and proper then:
\begin{enumerate}
\item 
$S/G_S$ is a differentiable manifold and the projection
$\pi\colon S\to S/G_S$ is a surjective submersion.
\item
There exists an unique closed form
$\tilde\omega\in\df^{k+1}(S/G_S)$
such that $\pi^*\tilde\omega=\omega$.

(Observe that, as ${\cal G}_S\subseteq{\cal K}_\omega$,
then $\tilde\omega$ is $1$-degenerate, in general).
\end{enumerate}
\label{redteo}
\end{teor}
\proof
The proof of the item 1 follows the same pattern as in the classical
theorems of reduction (see \cite{AM-78}, \cite{LM-87}, \cite{MW-rsms}).
For the second item, 
 from (\ref{GHpropo}),
it is easy to prove that $\omega$ is a $\pi$-basic form.
Thus the existence of $\tilde\omega$ is assured,
and it is obviously a closed form.  
The uniqueness is a consequence of
the fact that $\pi$ is a surjective submersion, and
hence $\pi^*$ is injective.
\qed

{\bf Remarks}:
\begin{itemize}
\item
If $\dim S<k+1$ then $\omega=0$ and the result is trivial.
Moreover $\dim S/G_S\geq k+1$, as 
${\cal G}_S\subset{\cal K}_\omega$.
\item
Observe that if we make the quotient $S/\ker\,\omega$
we obtain a reduced form $\tilde\omega\in\df^{k+1}(S/\ker\,\omega)$
which is nondegenerate (that is, multisymplectic), 
but then we have removed more degrees of freedom than 
the corresponding to the symmetries introduced by the group $G$.
\item 
In the case of reduction of
symplectic and presymplectic manifolds, 
if the action is {\sl Poissonian}
(that is, there exist a {\sl momentum map} $J\colon {\cal M}\to {\mathfrak g}^*$,
which is $Ad^*$-equivariant), there is a natural way of obtaining
momentum-type submanifolds: they are the level sets of
the momentum map, $J^{-1}(\mu)$, for every weakly regular value
$\mu\in{\mathfrak g}^*$. These level sets are the maximal integral submanifolds 
of the Pfaff system 
$\{\inn(\tilde\xi)\Omega=0\, , \, \forall\xi\in{\mathfrak g}\}$ \cite{EMR-99},
and all of them are optimal momentum-type submanifolds, since
in these cases $k=1$ and the equality in (\ref{dim}) holds
since it reduces to $s=n-m$.
\item
In the multisymplectic case, the momentum map ${\cal J}$ 
associated with a Poissonian action allows us to define 
an exterior differential system
which is generated by the forms 
$\{\inn(\tilde\xi)\Omega=\d{\cal J}^*(\xi)\, ,\, \xi\in{\mathfrak g}\}$,
and whose maximal integral submanifolds
are momentum-type submanifolds, wich are called the 
{\sl integral submanifolds of the momentum map} ${\cal J}$.
Then, theorem \ref{redteo} holds in this context.
\end{itemize}

\section{Noether invariants in Lagrangian field theory}

A Lagrangian field theory is characterized giving
a {\sl configuration bundle} $\pi\colon E\to M$,
where $M$ is a $k$-dimensional oriented manifold 
with volume form $\omega\in\df^k(M)$, and a
{\sl Lagrangian density} which is a
$\bar\pi^1$-semibasic $k$-form on $J^1\pi$,
where $\pi^1\colon J^1\pi\to E$ is the
jet bundle of local sections of $\pi$, and
$\bar\pi^1=\pi\circ\pi^1\colon J^1\pi \longrightarrow M$
gives another fiber bundle structure. We have that
$\Lag =\lag \bar\pi^{1^*}\omega$, where $\lag\in\Cinfty (J^1\pi)$
is the {\sl Lagrangian function} associated with $\Lag$ and $\omega$.
The {\sl Poincar\'e-Cartan forms} associated with $\Lag$, denoted
$\Theta_{\Lag}\in\df^{k}(J^1\pi)$ and
$\Omega_{\Lag}:= -\d\Theta_{\Lag}\in\df^{k+1}(J^1\pi)$,
are constructed using the canonical jet bundle
elements. $\ls$ is a {\sl Lagrangian system}, which
is {\sl regular} if $\Omega_{\Lag}$ is $1$-nondegenerate.
The Lagrangian problem consists in finding sections 
$\phi\in\Gamma(\pi)$ (the set of sections of $\pi$),
such that, if $j^1\phi$ denotes the canonical lifting of 
$\phi$ to $J^1\pi$, then
$(j^1\phi)^*\inn (X)\Omega_\Lag=0$, for every $X\in\vf (J^1\pi)$.
These are the Euler-Lagrange field equations.
(See \cite{EMR-96} for details).

Let $\ls$ be a regular Lagrangian system, and
$\Phi \colon G\times J^1\pi\to J^1\pi$ be an action of 
a Lie group $G$ on $J^1\pi$.
If $\Phi$ is an exact action  (hence strongly multisymplectic), and 
$\Phi_g^*\lag=\lag$, for every $g\in G$,
then $G$ is a {\sl symmetry group} of $\ls$. 
Therefore, as described above, a comomentum map exists for which 
the action is Coad-equivariant and Poissonian, and it is given by 
\beq
{\cal J}^*(\xi)=\inn (\tilde\xi )\Theta_\Lag
\ ,\ 
\mbox{\rm for every $\xi\in {\mathfrak g}$} \ .
\label{comom}
\eeq

But $\inn (\tilde\xi )\Theta_\Lag$ are just the
{\sl Noether invariants} associated with this symmetry group.
In fact, {\sl Noether's theorem} states that, for
every $\phi\in\Gamma(\pi)$ solution to the field equations, from 
$\Lie(\tilde\xi)\Theta_\Lag=\d\inn(\tilde\xi)\Theta_\Lag+
\inn(\tilde\xi)\d\Theta_\Lag=0$, for every $\xi\in{\mathfrak g}$,
we have
$$
0=(j^1\phi)^*\d\inn(\tilde\xi)\Theta_\Lag+
(j^1\phi)^*\inn(\tilde\xi)\Omega_\Lag=
(j^1\phi)^*\d\inn(\tilde\xi)\Theta_\Lag \ .
$$
Observe that
$(j^1\phi)^*\inn(\tilde\xi)\Omega_\Lag=0$
implies that
$\{\d\inn(\tilde\xi)\Theta_\Lag=\inn(\tilde\xi)\Omega_\Lag
\, ,\, \xi\in{\mathfrak g}\}=[\vf^k({\rm Im}\,j^1\phi)]'$, and
then, for every $\phi\in\Gamma(\pi)$ solution,
${\rm Im}\,j^1\phi$ are momentum-type submanifolds
of $J^1\pi$. So, we call them {\sl $G$-Noether-type submanifolds}.
All of this allows us to state:

\begin{prop}
If $N_G$ is a $G$-Noether-type submanifolds which contains
a Cauchy data submanifold $S$, and $\phi$ is a solution
to the field equations on it, then
${\rm Im}\,j^1\phi\subset N_G$.
\end{prop}

The remaining question is under what conditions
the reduction procedure described above can be applied to reduce
the field equations. Results in this way have been already obtained \cite{
FGR-2001,Sn-2004}.

Observe also that, if ${\cal J}^{'*}$ is another
comomentum map, then 
$$
\d{\cal J}^{'*}(\xi)=\inn(\tilde\xi )\Omega_\Lag=
\d\inn(\tilde\xi)\Theta_\Lag \ \Longrightarrow\
\d({\cal J}^{'*}(\xi)-\inn(\tilde\xi)\Theta_\Lag)=0 \ .
$$
Thus, it is obvious that the construction of 
$G$-Noether-type submanifolds depends only on the group action, 
and not on the choice of a comomentum map.
The relevant fact is the existence of the comomentum map
given by (\ref{comom}).

It is interesting to point out that, in the realm of
first-order Lagrangian field theories,
the existence of momentum-type
submanifolds is assured (the images of the sections
solution to the field equations). 
Nevertheless, we cannot assure
the existence of those manifolds verifying
the condition of maximal dimensionality
(i.e., for being maximal momentum-type submanifolds).

\section{Discussion and outlook}

Recently, a very generic scheme of reduction  
in the ambient of the so-called $k$-symplectic or
polysymplectic formulations of classical field theories
has been completed \cite{MRSV-2015},
eestablishing sufficient conditions in order to do this reduction possible.
In a previous paper \cite{RRSV-2011}, the relation between
the $k$-symplectic (polysymplectic) and the multisymplectic formalisms
(for certain kinds of multisymplectic manifolds) was studied.
Bearing in mind this relationship and the results in \cite{MRSV-2015},
we could study also the equivalence between the $k$-symplectic reduction
and a suitable reduction programme for the multisymplectic case.
This line of work will be the object of a future research.

Finally, another way of approaching the problem of the 
multisymplectic reduction could be using the so-called 
{\sl higher-Dirac structures}, since the reduction of these types of structures 
would generalize the multisimplectic and polysimplectic reduction, 
in the same way that the reduction of {\sl Dirac structures} 
encompasses other reduction procedures such as the symplectic 
(Marsden-Weinstein), presymplectic and Poisson cases
\cite{BMR-2017}.

\begin{small}

\section*{Acknowledgments}

We are grateful to Profs Juan Pablo Ortega for his valuable comments and suggestions 
on redution theory, to Prof. L. Alberto Ibort for the fruitful discussions on the construction of the multimoment um map, 
and to Prof. Roberto Rubio for the information about higher-Dirac structures and their role in this problem of reduction.
We thank Prof. Casey Blacker for his comments that have allowed us to correct an error in the definition of the bracket of Hamiltonian forms.
We acknowledge the financial support of 
{\sl Ministerio de Ciencia e Innovaci\'on} (Spain), projects
MTM2014--54855--P and MTM2015-69124--REDT.
and of
{\sl Generalitat de Catalunya}, project 2017-SGR932.
 We thank Mr. Jeff Palmer for his
assistance in preparing the English version of the manuscript.

\end{small}

\end{document}